\pdfoutput=1

\documentclass[11pt]{article}

\usepackage[final]{acl}

\usepackage{times}
\usepackage{latexsym}

\usepackage[T1]{fontenc}

\usepackage[utf8]{inputenc}

\usepackage{microtype}

\usepackage{inconsolata}

\usepackage{graphicx}
\usepackage{url}
\usepackage{booktabs}

\usepackage{tikz}
\usetikzlibrary{shapes, arrows, positioning}

\title{Evaluating the Impact of LLM-guided Reflection on Learning Outcomes with Interactive AI-Generated Educational Podcasts}

\author{
 \textbf{Vishnu Menon\textsuperscript{1}},
 \textbf{Andy Cherney\textsuperscript{1}},
 \textbf{Elizabeth B. Cloude\textsuperscript{2}},
 \textbf{Li Zhang\textsuperscript{1}},
 \textbf{Tiffany D. Do\textsuperscript{1}}
\\
\\
 \textsuperscript{1}Drexel University,
 \textsuperscript{2}Michigan State University
\\
 \small{
   \textbf{Correspondence:} \texttt{
   \{\href{mailto:vishnu.v.menon@drexel.edu}{vishnu.v.menon}|\href{mailto:harry.zhang@drexel.edu}{harry.zhang}|\href{mailto:tiffany.do@drexel.edu}{tiffany.do}\}@drexel.edu, \href{mailto:cloudeel@msu.edu}{cloudeel}@msu.edu
   }
 }
}

\begin{document}
\maketitle
\begin{abstract}
This study examined whether embedding LLM-guided reflection prompts in an interactive AI-generated podcast improved learning and user experience compared to a version without prompts. Thirty-six undergraduates participated, and while learning outcomes were similar across conditions, reflection prompts reduced perceived attractiveness, highlighting a call for more research on reflective interactivity design.
\end{abstract}

\section{Introduction}
What if educational content could not only speak to learners, but listen, adapt, interact, and assess learning processes -- like \textit{reflection} -- in real-time? As learners increasingly disengage from traditional materials like textbooks~\cite{textbook}, large language models (LLMs) offer new opportunities to deliver content in more engaging, interactive, and personalized formats, such as AI-generated podcasts~\cite{jin2025modeling}. Emerging tools like NotebookLM\footnote{\url{https://notebooklm.google/}} illustrate growing public interest in generative AI for learning.

Personalized learning with AI has been shown to support self-regulated learning by encouraging learners to plan, monitor, and evaluate their progress~\cite{shemshack2020systematic,molenaar2023measuring}. Prior work demonstrates that personalized AI-generated podcasts based on college textbooks (tailored to learners’ majors, interests, and instructional preferences) can enhance learning and enjoyment compared to both textbooks and non-personalized content~\cite{do2025paige}. However, most AI-generated podcasts remain \textit{passive}: learners can ask questions, but the system does not initiate interaction or assess learning to guide deeper engagement. This represents a missed opportunity, as structured interactivity has been shown to enhance engagement and active learning in other domains~\cite{Laban2022}. 

More importantly, reflection is a critical component of learning -- it helps learners draw meaningful and construct understanding in connection with learning goals. Embedding structured, reflection prompts into AI-generated podcasts could enhance engagement and learning, but may also disrupt learners' concentration and flow, possibly reducing their effectiveness. Design trade-offs remain unclear: when should reflection be prompted, and how can learners' responses be assessed in real-time with AI-generated podcasts?

We investigated these questions in a controlled experiment with a sample of 36 undergraduates, comparing two conditions: Reflection, where an AI-generated podcast periodically prompted learners to reflect and responded based on their input, and Standard, where no reflection prompts were prompted by the system. This study specifically investigates an AI-generated podcast featuring a single host, using two research questions: (1) Do interactive—in this case, meaning a model that can be freely interrupted, conversed with and asked questions—reflection prompts improve learning outcomes when incorporated into AI-generated podcasts compared to standard AI-generated podcasts? and, (2) Do interactive reflection prompts improve user experience when incorporated into AI-generated podcasts compared to standard AI-generated podcasts?

\section{Related Work}
Reflection is a key self-regulatory process that supports deeper learning and metacognitive awareness by promoting learners to contemplate their understanding and connect it with previous learning experiences.~\citet{mcalpine1999building} conceptualize reflection as a goal-driven process in which learners continuously integrate knowledge and action. Building on this, recent work has explored whether digital learning environments can scaffold reflection to improve learning outcomes.~\cite{cloude_carpenter} examined the impact of reflective prompts in a game-based learning environment with 120 adolescents. Learners received one of three types of reflection prompts during learning: progress planning, solution strategy, and different problem approaches. Findings showed that the quantity and quality of reflections influenced learning, but their effects varied depending on the learner's goals.

Carpenter et al. \citeyearpar{carpenter_gble} further investigated reflection quality in middle-school students, using a rubric to assess written responses on a 5-point scale ranging from non-reflective to highly reflective. Higher-quality reflections -- those including hypotheses, planning, and reasoning -- were more predictive of learning gains. However, these reflections were scored post hoc, underscoring a key limitation in current research: the inability to evaluate and respond to reflection \textit{in real time}. Cloude et al.~\citeyearpar{cloude_carpenter} highlight the lack of theoretical clarity around when and how to prompt reflection during learning, a gap this work aims to address by embedding structured, interactive prompts into AI-generated podcasts. In our system, an LLM-driven agent prompts learners to reflect and evaluates their spoken responses to guide real-time support.

While prior studies have explored personalized AI-generated podcasts for education, they have not addressed reflection. Do et al. \citeyearpar{do2025paige} compared generalized and personalized podcasts -- generated from textbook chapters using LLMs -- to traditional textbook reading on learning outcomes. Personalized podcasts, tailored to learners' majors and interests, improved enjoyment and learning outcomes. Their systems used a multi-stage generation pipeline with Gemini 1.5 Pro to convert textbook content into conversational podcast scripts, which were then synthesized using text-to-speech models.

Other work has explored AI-generated podcasts outside of education. Yahagi et al. \citeyearpar{yahagi2025paperwave} showed that transforming academic papers into AI-generated podcasts lowered barriers to engaging with academic literature. Similarly, Laban et al. \citeyearpar{Laban2022} examined AI-generated podcasts for news delivery and found that it enhanced enjoyment. However, unlike our system, these podcasts lacked interactive, reflective components and were not designed for structured learning contexts. Together, these lines of research inform our approach: we integrate structured reflection prompts -- adapted in real-time by an LLM -- into AI-generated podcasts to support engagement, reflection, and learning during listening.

\section{Interactive Podcast Architecture}
\begin{figure}[h]
\centering
\begin{tikzpicture}[
    node distance = 1cm,
    auto,
    thick,
    end/.style = {rectangle, text width=1.8cm, text centered, minimum height=0.7cm, font=\scriptsize},
    block/.style = {rectangle, draw, fill=blue!20, text width=1.8cm, text centered, rounded corners, minimum height=0.7cm, font=\scriptsize},
    final/.style = {diamond, draw, fill=yellow!20, text width=1.8cm, text centered, font=\scriptsize, inner sep=0.2pt},
    arrow/.style = {->, >=stealth}
]

\node [end] (std_start) {Begin};
\node [block, below of=std_start] (std_agent) {Agent Speaks};
\node [block, below of=std_agent, dashed] (std_user) {User Interrupts};
\node [end, below of=std_user] (std_finished) {End};

\draw [arrow] (std_start) -- (std_agent);
\draw [arrow] (std_agent) -- (std_user);
\draw [arrow] (std_user) -- (std_finished);

\draw [arrow] (std_user) -- ++(1.5,0) |- (std_agent);

\node [end, right=2cm of std_start] (ref_start) {Begin};
\node [block, below of=ref_start] (ref_agent) {Agent Speaks};
\node [block, below of=ref_agent, dashed] (ref_user) {User Interrupts};
\node [block, below of=ref_user] (ref_prompt) {Reflection Prompt};
\node [block, below of=ref_prompt, fill=yellow!20] (ref_response) {User Reflection};
\node [end, below of=ref_response] (ref_finished) {End};

\draw [arrow] (ref_start) -- (ref_agent);
\draw [arrow] (ref_agent) -- (ref_user);
\draw [arrow] (ref_user) -- (ref_prompt);
\draw [arrow] (ref_prompt) -- (ref_response);
\draw [arrow] (ref_response) -- (ref_finished);

\draw [arrow] (ref_user) -- ++(1.5,0) |- ([yshift=-1.75mm]ref_agent);
\draw [arrow] ([yshift=-1.75mm]ref_response.east) -- ++(0.8,0) |- ([yshift=1.75mm]ref_agent);
\draw [arrow] ([yshift=1.75mm]ref_response.east) -- ++(0.48,0) |- (ref_prompt);

\node [above=0.2cm of std_start, font=\small\bfseries] {Standard};
\node [above=0.2cm of ref_start, font=\small\bfseries] {Reflection};

\end{tikzpicture}
\caption{Standard vs Reflection Interaction Modes.}
\label{figure1}
\end{figure}
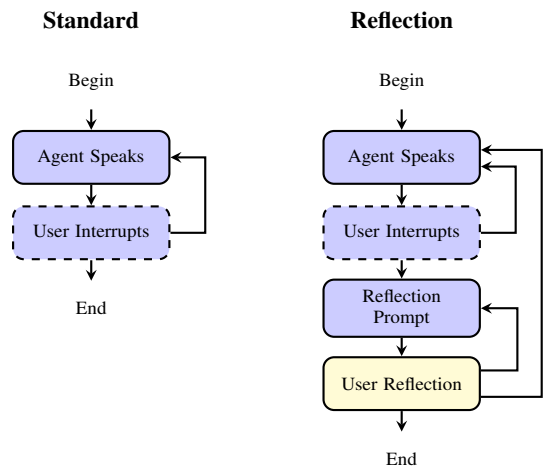

The technical implementation of our AI-generated podcast system involves ingesting textbook content and delivering an interactive podcast on demand. The system supports two modes of interaction (Figure \ref{figure1}):

\begin{itemize}
  \item \textit{Standard}: The system delivers audio content continuously from the textbook and allows learners to interrupt at any time with questions or comments. This interaction style is similar to existing consumer podcast systems, such as NotebookLM’s Interactive mode. 
  \item \textit{Reflection}: The system delivers audio content and incorporates structured reflection prompts, periodically pausing after key concepts are introduced, and requiring the learner to demonstrate understanding of the content before continuing in their spoken reflection.
\end{itemize}

\begin{figure*}[h!]
  \centering
  \includegraphics[width=5in]{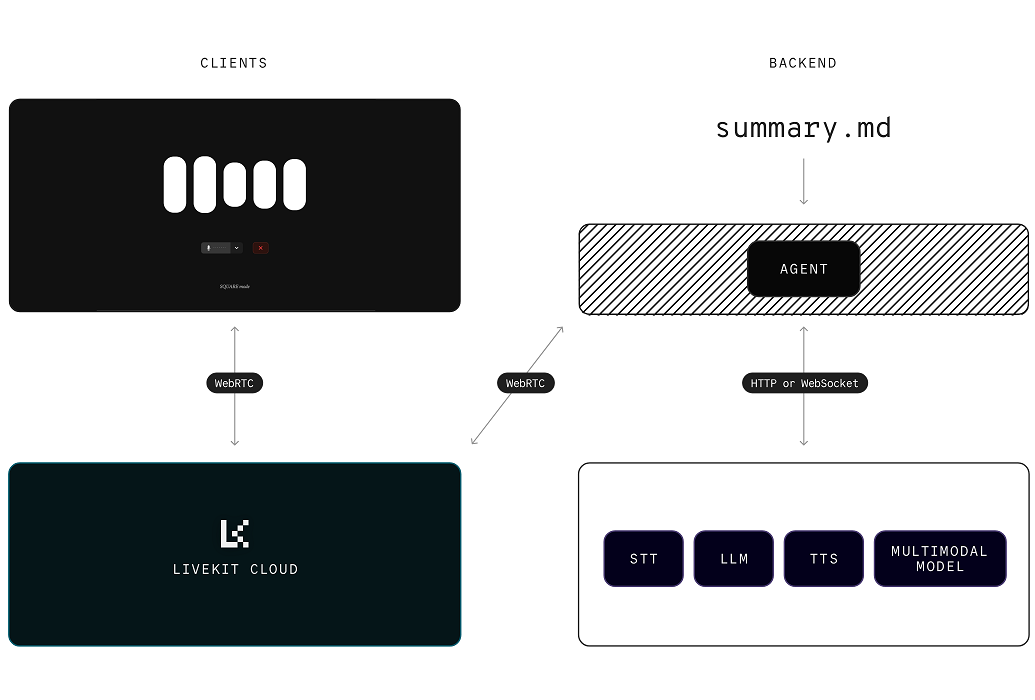}
  \caption{A diagram of our system, adapted from LiveKit's Agents Overview.}
  \label{fig:diagram}
\end{figure*}

The system architecture is shown in Figure \ref{fig:diagram}. The system consists of a Python backend for content generation, which hosts a LiveKit\footnote{\url{https://livekit.io/}} room and creates an agent for speech synthesis. LiveKit is a platform for building AI-voice applications that can interact with users over the web.

The frontend is built with React, Next.js, and TailwindCSS, and connects to the backend to enable real-time communication with the system. Based on these components, our system achieves an average response latency of 300ms \cite{dsa2024openai}, closely matching the pace of natural human conversation \cite{turn_taking}. This low-latency architecture, which we further detail below, enables controlled comparisons between the two interaction paradigms. To support reproducibility, we have open-sourced the system; the code repository is available on GitHub\footnote{\url{https://github.com/DU-DIVALab/tutorflow}}.

\subsection{Structured Summary}
The system first ingests Chapter 1, Section 1.1 of OpenStax's textbook \textit{Introduction to Philosophy} \cite{smith_2022}. Then, using GPT-4 Turbo, it converts academic text into a structural and summarized skeleton for the podcast, following the summarization procedure outlined by Laban et al. \citeyearpar{Laban2022}. We found that using Laban et al.'s structured summarization approach to generate podcasts addressed several challenges, such as material omission. When generating podcasts directly from source text, we found that the model often omitted important material and produced outputs constrained by its context window, regardless of input length. This created an artificial ceiling on content length and limited scalability for longer educational materials. By using a structured summary, where each section corresponds to a paragraph from the original source, we were able to generate each segment independently, ensuring content coverage and improving quality, similar to skeleton-of-thought~\cite{ning2023skeleton}. The structured summary also improves interpretability, providing transparency into the generation process and facilitating easier debugging. It serves as a reference to track content coverage during the learner-facing conversation.

\subsection{Podcast Generation} 
The structured summary is first divided into segments according to its outline and then processed by GPT-4o-mini. Each segment is used to generate corresponding portions of the podcast. Using OpenAI’s GPT-4o text-to-speech (TTS) model with the \textit{Alloy} voice, the podcast is synthesized as natural-sounding speech, incorporating appropriate pacing and intonation based on the skeleton structure.

\subsection{User Interaction and Reflection} 
We serve the content to the learner differently depending on their current interaction context, tracked via state machine. In the Reflection mode, it monitors learner responses to assess their knowledge of the topic and prompt reflections by asking: ``So, what is the most important thing you've learned so far?'' at the end of each section, following similar prompts by~\cite{cloude_carpenter}. After the learner responds to the reflection prompt, we use a one-shot evaluation to determine whether the response is suitable using a binary assessment (1 = demonstrates understanding, 0 = does not demonstrate understanding). The learner's answer is considered satisfactory if the prompt ``demonstrates awareness of their own knowledge'' This is to ensure the learner cannot proceed with the audio session simply by restating a keyword the model used back to it, as previous work showed that domain-specific words in reflections were not indicative of the quality of reflection, but rather reflective depth \cite{cloude_carpenter}. 

We employed in-context learning \cite{dong2022survey} using examples in the prompt to guide the agent's judgment. An example was if a learner is listening to content about Confucius, and they respond to a reflection prompt as ``Confucius'' to the model, this—while technically not incorrect, we guided learners to provide a more, detailed response to demonstrate synthesis of their knowledge to demonstrate a suitable reflection. For example, a response to a prompt with ``Confucius' teachings would be considered patriarchal by modern standards'' demonstrates a learner's understanding by combining a facet of what the content learned with how they contextualize the subject to present day. 

It is important to note that the reflection prompts are different from quizzes. The learner does not need to mention everything they learned about the topic, only by demonstrating their ability to synthesize new understanding, the model deems their engagement and reflection on the material. The binary satisfactory/unsatisfactory classification acts as an elegant gate to guiding learner's progress in real-time to capture reflective depth, as opposed to relying on keyword matching, while avoiding the complexity of multi-dimensional rubrics. In the Standard mode, our system continuously listens for interruptions but otherwise continues speaking until one occurs, and does not prompt the learner to reflect. During learner interactions, the system uses Deepgram for learner speech transcription, and Silero VAD \footnote{\url{https://github.com/snakers4/silero-vad}} to detect when learners were speaking. Additionally, a fine-tuned SmolLM v2 model \cite{allal2025smollm2smolgoesbig} predicts speech boundaries to support smooth turn-taking. 

\subsection{Podcast Interface} 
As shown in Figure \ref{fig:diagram}, the web application displays a decorative wave, an abstract animated visualization of the generated speech that animates based on the volume and cadence of the AI podcaster's voice. When the podcaster is silent, the wave appears as a flat line of dots. Learners begin the session with their microphone automatically turned on after granting permission through their browser.

\section{Methods}
\subsection{Sample}
To build on the methods used by Do et al. \citeyearpar{do2025paige}, we designed our study as an extension of their work and used the same source material and measurements. This study was approved by an Institutional Review Board and a total of 36 ($n$=36; 42\% female) college students enrolled at universities in the United States were recruited through the Prolific online marketplace. Participants were pre-screened for English fluency and minimal prior knowledge of the subject (Introductory Philosophy). We also screened participants for technical requirements to ensure they had a working microphone and speaker for audio. One participant was excluded from our analysis due  to adversarial responses to reflection prompts (e.g., ``I hate bots and I hate them in the work place [sic]''). Due to the added length introduced by the interaction in the Reflection condition, we limited the scope to a single textbook, Introduction to Philosophy (Chapter 1) \cite{smith_2022}, and focused only on one subsection (Chapter 1.1), to ensure a manageable session duration while maintaining consistency with the original study design.

\subsection{Procedure}
The study consisted of a single 40-minute remote session. Before the session, participants were randomly assigned to the 1) Reflection or 2) Standard condition. Next, participants completed a brief demographic survey and were informed they would interact with an AI-generated podcast to learn about philosophy, after which we collected informed consent. Participants were then directed to a web application (described in Section 3) and guided through an interactive AI-generated podcast lasting approximately 15 minutes. Upon completion, the agent provided a verbal code and displayed a popup in the browser, enabling participants to proceed. They were redirected to a survey, where they completed the learning outcomes test and the User Experience Questionnaire (UEQ). Participants were compensated with \$10 USD after finishing the study.

\subsection{Dependent Variables}
\subsubsection{Learning Outcomes}
Learning was assessed using items from the post-chapter test bank from the OpenStax textbook\footnote{We do not provide the questions and answers for the knowledge retention questionnaires due to OpenStax policy. Verified educators from academic institutions may access test banks directly through OpenStax.}. The assessment items included seven multiple-choice questions from the Section 1.1 test bank. Due to the small number of multiple-choice questions for the single subsection, we included adapted three open-response items into multiple-choice questions, known as “Review Questions” \cite{smith_2022}, resulting in a total of 10 questions (see adapted items in Appendix \ref{sec:appendixB}). Statistical analysis revealed no significant score differences between the original and supplemental items ($ps>.05$).

\subsubsection{User Experience}
User experience was measured using the User Experience Questionnaire (UEQ)~\cite{laugwitz2008construction} immediately after the audio session, specifically the \textit{Attractiveness} and \textit{Stimulation} subscales (see Appendix \ref{sec:appendixA}). These subscales gauge the overall appeal and engagement of user's experience during the interaction, respectively. The UEQ employs a 7-point anchored Likert scale using adjective pairs such as “Annoying–Enjoyable” for \textit{Attractiveness} and “Demotivating–Motivating” for \textit{Stimulation}. We measured \textit{Attractiveness} and \textit{Stimulation} by averaging item scores within each subscale.

\section{Results}
Before conducting statistical analysis, we assessed whether our data adhered to a normal distribution using histograms and the D'Agostino-Pearson test, which suggested that our data were normally distributed across all variables ($K^2=2.56, p=.28$). 

\subsection{Research Question 1}
To address our first research question,  do interactive reflection prompts improve learning outcomes when incorporated into AI-generated podcasts compared to standard AI-generated podcasts, we calculated a two-sample $t$-test to compare whether there were differences in learning outcomes between Reflection and Standard conditions. The results suggested that there were no differences in learning outcomes between Reflection and Standard conditions, $t(34) = 0.89, p = 0.38, D = 0.29$.

\subsection{Research Question 2}
\begin{figure}[h]
  \centering
  \includegraphics[width=2.6in]{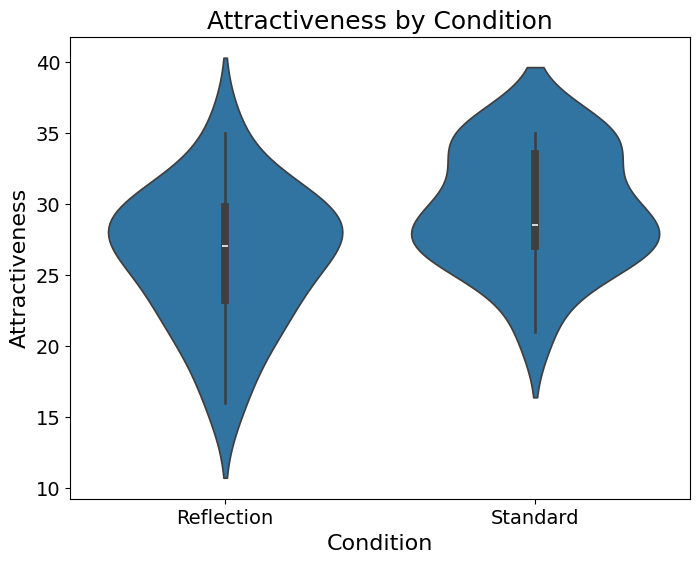}
  \caption{Attractiveness ratings across conditions.}
  \label{fig:attractive}
\end{figure}

To address our second research question, do interactive reflection prompts improve user experience when incorporated into AI-generated podcasts compared to standard AI-generated podcasts, we calculated 2 separate two-sample $t$-tests to compare whether there were differences in user experience subscales: attractiveness and simulation. The results suggested there was a significant difference in \textit{Attractiveness}, $t(34) = 2.26, p = 0.03, D = 0.75$, where the Standard condition rated the experience more favorably than the Reflection condition (Figure~\ref{fig:attractive}). Conversely, there were no significant differences in \textit{Stimulation}, $t(34) = 1.31, p = 0.20, D = 0.44$, between the Standard and Reflection conditions. Descriptive statistics are in Table \ref{table:stats}.

\begin{table}[h!]\small
\caption{Dependent variable descriptive statistics by condition.}
\begin{tabular}{@{}lccc@{}}
\toprule
 & \textbf{Learning} & \textbf{Attractiveness} & \textbf{Stimulation} \\ 
 \midrule
\textbf{Reflection} & 5.89 (1.94) & 26.22 (4.58) & 21.22 (3.68) \\
\textbf{Standard} & 6.50 (2.06) & 29.56 (4.00) & 23.17 (4.87) \\ \bottomrule
\end{tabular}
\label{table:stats}
\textit{Note.} Means and (standard deviations) are provided.
\end{table}

\section{Discussion}
Personalized learning via interactions and reflection prompts are both recognized as valuable tools to enhance engagement and active learning  \cite{interactive2019} \cite{reflection2023}. We implemented reflection prompts guided by~\cite{mcalpine1999building}'s model of reflection and empirical literature suggesting that deeper reflection enhances learning, as supported by prior research that included no evaluation of responses in real-time~\cite{cloude_carpenter,carpenter_gble}. To build on this, our system applied a one-shot evaluation to judge learners' understanding based on their spoken reflections. Our study sought to explore whether integrating reflection prompts within an interactive, AI-generated educational podcast would improve user experience and learning outcomes compared to a standard, interactive AI-generated podcast.

Our first research question revealed that interactive, AI-generated podcasts with reflection prompts did not significantly improve learning outcomes compared to those without reflection prompts. This result contrasts with previous research, which found that reflection enhanced learning outcomes in game-based environments \cite{cloude_carpenter,carpenter_gble}. One explanation could be the nature of the prompt, which asked learners to recall factually relevant information rather than promoting reflection in relation to learning goals or planning, which encourages more thorough reflection and which is required by a game-based environment. Another possibility is that while reflection may be a useful tool for AI-generated podcasts, simply relying on the LLM to guide the reflection process without accounting for the learning goals or knowledge state of the learner may not be effective for promoting reflection with interactive AI-generated podcasts.

In our second research question, we found that the reflection prompts with an interactive, AI-generated podcast significantly reduced \textit{Attractiveness} ratings compared to the standard AI-generated podcast condition. This indicated that the interactive elements in the Reflection condition may have disrupted the learners' flow and enjoyment during audio-based learning. This is different than previous research \cite{Wang2025}, which found that perceived interactivity and reflection boosted enjoyment and facilitated more active learning. The lack of significant differences for \textit{Stimulation} suggests that the reflection intervention, despite its theoretical foundation in enhancing learning through reflective scaffolding~\cite{mcalpine1999building}, did not measurably improve user experience or learning outcomes in our study. Effective podcast-based reflections with LLMs likely require more detailed scaffolding, such as fine-tuning the model to provide automatic, tailored feedback that is based on learners' individual goals and current knowledge state. Future work should focus on developing stronger guidance methods to support reflection. This addresses a limitation identified by Do et al., who reported that participants desired "opportunities for active engagement" with AI-generated podcasts \citeyearpar{do2025paige}, and suggests that while the specific implementation of reflection prompts may have detracted from the user experience, the general concept of interactive learning remains appealing to learners. The challenge appears to be finding the right balance between maintaining content flow and providing meaningful opportunities for reflection that effectively support learning.

\subsection{Limitations}
This study has important limitations to consider. First, our sample size of 36, which may limit the statistical power and generalizability of our results. Moreover, the focused scope of the content being taught (one section of a chapter) may not fully represent how reflection impacts learning across different subjects. Furthermore, our reflection responses were evaluated using a binary metric (understood/not understood) rather than evaluating the depth of reflection. This methodological constraint, though appropriate for our specific learning context, may have reduced the potential effectiveness of the reflection intervention compared to more elaborate implementations with different types of prompts. There are likely degrees to understanding which learning tools often fail to capture. Perhaps a human learner may feel more inclined to skip content they aren not understanding only to return to it later. Finally, the learner was exposed to the content for only 15 minutes, which may have reduced their learning and reflection due to the short nature of that task.

\subsection{Future Work}
Future research should explore alternative approaches for incorporating reflection and interaction in AI-generated podcasts. Developing adaptive reflection systems using LLMs that dynamically adjust based on learner engagement and metacognition would be a promising direction. Future work should investigate the use of LLMs for more fine-grained grading approaches for reflection quality, moving beyond binary assessments to evaluate responses with greater nuance. Additionally, investigating whether multi-modal data could better inform interaction and how to prompt reflections (e.g., eye movements, physiology, facial expressions, prior reflection quality, etc.) to enhance understanding.

\bibliography{custom}

\appendix

\newpage 
\section{User Experience Questionnaire}
\label{sec:appendixA}

All items were assessed on a 7-point scale, with the terms as anchors, adapted from \cite{laugwitz2008construction}. 

\noindent \textbf{Attractiveness} \\
\noindent Annoying – Enjoyable \\
\noindent Bad – Good \\
\noindent Unlikeable – Pleasing \\
\noindent Unpleasant – Pleasant \\
\noindent Unattractive – Attractive \\
\noindent Unfriendly – Friendly \\
\noindent \textbf{Stimulation} \\
\noindent Inferior – Valuable \\ 
\noindent Boring – Exciting \\
\noindent Not interesting – Interesting \\
\noindent Demotivating – Motivating

\section{Review Questions}
\label{sec:appendixB}
We adapted three additional questions from the free-response items in \cite{smith_2022} into multiple-choice format, alongside the chapter questions. 

\begin{flushleft}
\textbf{1. What characteristics are essential for being identified as a “sage”?} \\
a) Upholding social norms and exercising political power \\
b) Seeking profound understanding through critical inquiry and providing foundational insights \\
c) Mastering persuasive rhetoric and accumulating significant wealth \\
d) Adhering to religious doctrines and conducting spiritual rituals \\

\textbf{2. What does it mean for philosophy to “have an eye on the whole”?} \\
a) Rejection of traditional narratives through empirical investigation \\
b) Fusion of mystical beliefs with systematic logical analysis \\
c) Skeptical inquiry into established wisdom and foundational explanations of reality \\
d) Emphasis on practical skills for societal and technological advancement \\

\textbf{3. Which philosopher held that moral behavior and social harmony were linked to the natural order?} \\
a) Confucius \\
b) Pythagoras \\
c) Thales \\
d) Yajnavalkya

\end{flushleft}

\end{document}